{
\documentstyle[revtex,preprint]{aps}
\begin{document}
\preprint{BNL \#47712}
\begin{title}
Gauge Dependence of Effective \\ Quark Mass
and Matrix Elements in \\
Gaugefixed Large $N$ Strong Coupling Lattice QCD
\label{strongcouplechapt}
\end{title}
\author{Ken Yee\cite{newadd}}
\begin{instit}
Physics Theory Group, Brookhaven National Laboratory, Upton, NY  11973
\end{instit}
\begin{abstract}
{\baselineskip19pt
	In conjunction with recent numerical
\hbox{$\lambda~\partial_0 A_0 + \nabla\cdot\vec{A} =0$}
``$\lambda$-gauge'' results reported
in a companion paper, we construct
an $N\to\infty$ Wilson loop picture of $\lambda$-gaugefixing
in which (I)the $\lambda$-gauge expectation value of a
link chain $C$ is the weighted
sum over Wilson loops made by joining to $C$ all
selfavoiding chains $\widetilde{C}$ closing $C$.  (II)Weights
$A_{\widetilde{C}}$, containing all the $\lambda$-dependence,
are given by the $\beta=0$
$\lambda$-gauge expectation value
of $\widetilde{C}$.  (III)$A_{\widetilde{C}}$
equals path-products of coefficients from the
trace expansion of the gaugefixing Boltzmann weight.  From
(II) and (III) we deduce formulas for $\beta =0$
quark matrix elements.  We find that
$M_q^{(\lambda)}$ decreases
with increasing $\lambda$; the quark propagator
dispersion relation is not covariant
when $\lambda\ne 1$; and $\Delta I=1/2$ matching
coefficients are $\lambda$-independent.  These strong
coupling features are qualitatively
consistent with numerical $\beta=5.7$ and $6.0$ results
briefly described here for comparison purposes but mainly
presented in a companion paper.

}
\end{abstract}
\newpage
\baselineskip21pt

\section{Motivation and Results}

	Traditionally gaugefixing is done only in weak
coupling perturbation theory, where it is needed to
define perturbative quark and gluon propagators.  In lattice QCD,
gaugefixing is unnecessary for computing gauge invariant
correlation functions.   However, since local
gauge symmetry cannot break spontaneously \cite{elitzer}
the only way
to see nonvanishing gauge variant correlation functions is by
gaugefixing.  Accordingly, lattice
gaugefixing has drawn considerable attention
in recent years~\cite{ogilvie}.  As described in our companion
paper~\cite{bs}, gluon~\cite{mand1},
quark~\cite{cap1,tall1} and photon propagators~\cite{all},
effective masses, and
wavefunctions~\cite{labrenzcom} have been
studied in special cases of ``$\lambda$-gauges''
\begin{equation}\label{eq:lambda}
\lambda  \partial_0 A^0  + \nabla\cdot \vec{A} = 0 .
\end{equation}
The numerical work has prompted analytical and
computational studies
of longitudinal \cite{kyee,thesis} and
topological \cite{parrinello} gaugefixing ambiguities and
their effects on gluon, quark \cite{bs} and
photon \cite{all} correlation functions and operator
product expansion coefficients determined from gauge
covariant matching conditions~\cite{kyee,thesis}.

 	Nonperturbative gaugefixing is a complex
subject.  Since quarks are
confined $M_q$ may (or may not)
depend on gauge.  For example~\cite{kyee}, compare the
exactly solvable Schwinger model in Coulomb gauge to
covariant
gauges parametrized by gauge parameter $\xi$.  (We
were not able to solve the
model in $\lambda$-gauges.)  While the actual
situation in dimension $D=4$ QCD and lattice QCD may be
arguably different, it is helpful to have a litmus test
for discarding broad arguments (``Mass is gauge invariant in
perturbation theory; hence quark
mass is gauge invariant.'') which do not distinguish between
the Schwinger model and other gauge theories.  In
the Schwinger model, quarks are
confined but the photon
is physical and has a mass from the $U_A(1)$ anomaly.  We define
``effective mass'' as the inverse correlation length of the
zero momentum propagator.  While
the photon has a gauge invariant effective mass (equal to
its physical mass),
the effective quark mass varies with gauge
parameter $\xi$.  Coulomb gauge---the unitary gauge for
the Schwinger model---has no unphysical modes and the
quark propagator is the {\it physical\/}
amplitude for quark propagation.  Due to dielectric
breakdown of the vacuum~\cite{susskind1}, the
quark propagator---the amplitude to have just one
quark at $x\ne 0$ starting with an $x=0$ quark---vanishes
and the effective Coulomb gauge quark mass diverges.
In covariant gauges,  the presence of unphysical modes
(or, alternatively, Gupta-Bleuler physical-state conditions
on the Hilbert space) ruins this physical
interpretation of the quark propagator.  Covariant gauge
quark propagators are not amplitudes for quark propagation
and do not vanish despite confinement.

	As reported in our companion paper~\cite{bs},
effective quark and gluon masses $M_q$ and
$M_g$ were evaluated on $\beta=5.7$ and $6.0$
quenched $D=4$, color $N=3$ Wilson lattices in $\lambda$-gauges
by matching gaugefixed $\vec{p}=0$ propagators at
large $t_E$ to the free particle ansatz
\begin{equation}\label{ansatz}
\lim_{t_E\to\infty}\sum_{\vec{x}}{\rm e\/}^{i\vec{p}\cdot\vec{x}}
\langle V_x\psi_x\overline{\psi}_0
V_0^{\dag} \rangle = Z_q^{(\lambda)}
\Bigl({M_q^{(\lambda )}
+i{\rlap{$\slash$}{p}}\over
2E_q^{(\lambda)}} \Bigr) {\rm e\/}^{- E_q^{(\lambda)}
\vert t_E\vert}~,
\end{equation}
where \hbox{$E_q^{(\lambda )}(\vec{p}=0)
\equiv M_q^{(\lambda )}$} and
$Z_q^{\rm free\/} = 1$.  The ``$(\lambda)$'' superscript
anticipates $\lambda$-dependence, although sometimes we will
omit it for brevity.  The role of background gauge field
gauge transformations $V_x$ will be
explained shortly.  The idea of monitoring chiral symmetry
with quark masses goes back to
the Gross-Neveu model \cite{neveu}, where
the flavor $N_f\to\infty$ effective quark mass
\hbox{$M_q   =  M_c +  \pi  m_q +  {\cal O\/}(m_q^2)$}
is a continuous, increasing function of
bare mass $m_q$.  Similarly, matching
the vacuum expectation value of the operator product expansion
of \hbox{${\psi_x} {\overline{\psi}_0}$}
in massless QCD to a free fermion propagator yields for
$N=3$, flavor $N_f =3$ and Landau gauge \cite{politzer}
\begin{equation}\label{eq:mpolitzery}
\lim_{\mu^2 > -p^2\to \infty} M_q
(\mu ,p^2 )~\sim~ {4  g^2 (\mu_o )\over p^2}
 \Bigl( { g (\mu)\over g(\mu_o )}\Bigr)^{8/9}
\langle \bigl(\overline{\psi}_0\psi_0\bigr)_{\vert\mu} \rangle .
\end{equation}

	Numerical fits in lattice QCD to
\hbox{$M^{(\lambda )}_q
=  b^{(\lambda)}  M_\pi^{2}  + M^{(\lambda )}_c$}, motivated
by the CPTh relation
$M_\pi^2 \propto m_q$ between pion and
current quark mass \cite{cap1},
yield $b^{(1)}   \sim 2.7(.3)\times 10^{-4}/MeV$ and
$M^{(1)}_c   \sim 350(40)MeV$ at $\beta\sim 6.0$.  At
both $\beta=5.7$ and
$6.0$ effective quark and gluon masses
decrease as $\lambda$ grows \cite{tall1} so
that, roughly, $M^{(2)}_q /M^{(1)}_q \sim .9$ and
$M^{(25)}_q /M^{(1)}_q\sim .75$, all plus or minus $\sim 15\%$
jackknife errors.

	In this paper, we put forth a
$\lambda$-gauge $\beta=0$, color $N\to\infty$
solution of
lattice QCD with quenched Wilson fermions
in an infinite volume lattice.  Lattice gaugefixing
is implemented by a lattice Fadeev-Popov method; we
couple the links to a quenched Higgs gaugefixing
field $\{V_x\}$ in the fundamental
representation of $SU(N)$~\cite{ogilvie}.  Gluons
do not propagate
at $\beta =0$, where link fields
oscillate randomly.  Let us focus momentarily on quark
propagators.  $\lambda$-gauge
quark propagators are $V_x\psi_x$ ``meson'' propagators
in this formulation---as written
in~(\ref{ansatz}).  An expression for $V_x\psi_x$ meson
propagators follows from hopping
expanding the Higgs and quark fields,
doing the $\beta=0$ link integrals---which project out all but
zero area Wilson loops---and
resumming the hopping expansions.  However, as described
in Section~\ref{trexpand}, we are unable
to resum graphs where Higgs paths
recur (Recurrence and other such notions
are defined in Section~\ref{sec:defs}.)
since the hopping expansion weight of such paths differ
from same-pathlength nonrecurrent
graphs.  To get around this we resort to a
``trace orthogonality approximation,'' which does not
differentiate between
recurrent and nonrecurrent Higgs paths.  This approximation
is tantamount to taking the $N\to\infty$ limit
{\it before\/} resumming because, as we show
in Section~\ref{trexpand}, there
is no difference between recurrent and nonrecurrent paths
in the brutally truncated
$N\to\infty$ limit.  Taking $N\to\infty$ before
resummation is an approximation because
the hopping expansion is (apparently)
not absolutely convergent at infinite $N$.

	In the
trace orthogonality approximation, only
selfavoiding quark paths contribute to the $\beta =0$
quark propagator because the $\{V_x\}$, being quenched,
dress only quark paths without
internal loops.  Hence nonselfavoiding quark paths are suppressed
by infinite string tension at $\beta=0$.  However, since are unable to
(re)sum over only selfavoiding quark paths, we make
an additional approximation and sum over
all (for technical reasons) nonbacktracking
quark paths.  When we do this we find, as
shown in Section~\ref{basics},
that the $\beta=0$, $r=0$, $N\to\infty$
zero momentum quark propagator pole
$M_q^{(\lambda )}$ is analytic and linear
in $m_q$ as $m_q\to 0$.  Expanding
\begin{mathletters}
\begin{equation}\label{eq:linep}
M_q^{(\lambda)} = M_c^{(\lambda )} +  B^{(\lambda )}
m_q + {\cal O\/}(m_q^2)
\end{equation}
yields
\begin{equation}\label{eq:mcd0p}
M_c^{(\lambda )} ={\sqrt{2D-1}\over 2(32D-7)\lambda }
\cases{ 11+14D-25\lambda^2 & $\lambda \le {1\over 2}$;\cr
{-25+200\lambda + 224(D-1)\lambda^2\over 4 (4\lambda
-1)} &$\lambda\ge {1\over 2}$;\cr}
\end{equation}
\begin{equation}\label{eq:bl0p}
B^{(\lambda)} =\cases{~~~~
{499-1474D + 1600 D^2 + (1376D-751)\lambda^2 \over
2\lambda (32D-7)^2 } & $\lambda \le {1\over 2}$;\cr
{ -751 + 1376 D + 6008\lambda - 11008 D\lambda + (-4032 -1568D
+ 25600D^2 )\lambda^2 \over 8 (32D-7)^2 \lambda (4\lambda -1) }
&$\lambda \ge {1\over 2}$.\cr}
\end{equation}
\end{mathletters}
Note that the $\lambda$-dependence of $M_c^{(\lambda)}$
is qualitatively the same whether
$D\to\infty$ (where the number of recurrences and
selfintersections are negligible) or $D=4$.  This
suggests that recurrences and selfintersections are
not responsible for $\lambda$-dependence.  Therefore we
generally quote formulas for all $D$ when we might be safer
with their $D\to\infty$ limits.  In any case,
we do not find any
qualitative difference between the $\lambda$-dependence
of $M_q^{(\lambda )}$ at finite and infinite $D$ in our
approximation scheme.
\vspace{3.5in}
\figure{$\lambda$-dependence comparison of
numerical quark masses to the
$\beta=0$, $N\to\infty$ Formula~(\ref{eq:mcd0p})
evaluated at $D=4$.  The numerical masses are
rescaled so that at $\lambda =1$ all
data points are normalized to
$M_c^{(1)}$.  Ref.~\cite{bs} provides details
of the numerical simulation. \label{figs2}}

	As depicted in Fig.~\ref{figs2}, Eq.~(\ref{eq:mcd0p})
mimics the $\lambda$-dependence of
the $\beta=5.7$ and $6.0$ numerical
data.  Most of the change occurs between
$\lambda=1$ and
$\lambda=2$, and $M_q^{(\lambda)}$ stabilizes to a
nonzero value as $\lambda\to\infty$.  This qualitative
agreement between strong coupling and numerical behavior
helps give confidence that the
numerical $\lambda$-dependence of $M_q$
is not a finite volume artifact or due to
details of how $M_q$ is extracted in the numerical simulations.

	For technical reasons we
compute the quark propagator dispersion relation not at
$r=0$ but at $r=1$, where it is
\begin{mathletters}
\begin{equation}\label{dispersion1p}
{E_q^{(\lambda )}}^2 = {M_q^{(\lambda )}}^2
+ \sum_{i=1}^{D-1} \overline{g}_i^2 ~ \vec{p}_i^2 + {\cal O\/}(a^3)~,
{}~~~~ E_q^{(\lambda)} \equiv \pm i p_0~,
\end{equation}
\begin{equation}\label{gidefp}
\overline{g}_i ( \lambda)\equiv\cases{ {3\over 4\lambda} &$\lambda
\le {1\over 2}$;\cr {3\lambda\over 4\lambda -1}&$\lambda
\ge {1\over 2}$.}
\end{equation}
\end{mathletters}
Since $\overline{g}_i(\lambda\ne 1)\ne 1$, the quark dispersion
relation is not covariant and, hence, the propagator is not
free particle-like except in Landau
gauge.  $\overline{g}_i$ drops out
if $\vec{p}=0$, where
the quark propagator is indistinguishable from the free particle
propagator.  (However $M_q^{(\lambda )}$ remains
$\lambda$-dependent.)

	As described in Section~\ref{sec:matchingsc},
matching coefficients for $\overline{s}d$ subtraction
of $\Delta I=1/2$ Rule operators in the $\beta=0$, $N\to\infty$
limit assuming trace orthogonality are given by
\begin{equation}\label{alphab0valp}
\alpha_{{O_\pm} [\Gamma_1\Gamma_2]} ~=~
\bigl(\delta_{\Gamma_1\Gamma_2,SS} \mp {1\over 2^{D/2} N}
f^{SS}_{\Gamma_1\Gamma_2} \bigr)~\langle \overline{\psi}_0
\psi_0\rangle
\end{equation}
where Fierz coefficients $f^{\Gamma_3\Gamma_4}_{\Gamma_1\Gamma_2}$
are given in Eq.~(\ref{fierzratsp}).  Since
\hbox{$\langle\overline{\psi}_0
\psi_0\rangle$} is gauge invariant these $\beta=0$ matching
coefficients,
which are directly related to
physical continuum decay rates, are gauge
invariant.  At $\beta =6.0$, $\alpha_{{O_\pm}
[\Gamma_1\Gamma_2]}$ also seem to be $\lambda$-independent
up to statistical errors~\cite{bs}.

\section{Lattice Gaugefixing}
\label{sec:gaugefixing}

	As exemplified by several models---notably QED$_{3+1}$---the
thermodynamical limit of strong coupling lattice gauge theories
may correspond to different field theories than
the $\beta\to \infty$ ones.  In particular, proving gauge
dependence of quark mass in the lattice Schwinger model
would not reveal much about quark mass in the continuum
Schwinger model since the lattice Schwinger model has a phase
transition in $\beta$~\cite{vertexmod}.  (This transition, if
the critical point is unique,
doesn't ruin confinement since Schwinger model quarks
are confined at both weak and strong coupling.)  Analogously, the
following $\beta=0$, $N\to\infty$ solution should be
viewed as a toy model and not something necessarily related to QCD.

	The $\lambda$-gauge lattice expectation value of
a lattice operator $\Theta$ is
\begin{mathletters}
\begin{equation}\label{expectdef}
\langle {\Theta} \rangle ~\equiv~ \Bigl[~\bigl[~[
{\Theta} ]_f \bigr]_v \Bigr]_u
\end{equation}
where~\cite{ogilvie}
\begin{equation}
[\Theta]_{\theta}~\equiv~
z_\theta^{-1}  \int [d\theta ]  {\rm e\/}^{-S^\theta} \Theta~,
{}~~z_\theta \equiv
\int [d\theta' ]  {\rm e\/}^{-S^{\theta '}}   ,
{}~~ \theta\in\{f,v,u\},
\end{equation}
\begin{equation}
S^f \equiv \sum_{x,y} \overline{\psi}_x\Bigl[ \delta_{x,y}  +
K_B \sum_{n}
(r-\rlap{$\slash$}{n} ) U_{x,n} \delta_{y,x+\hat{n}}\Bigr]\psi_y ~,
\end{equation}
\begin{equation}\label{svdef}
S^{v} [\Omega ] \equiv -\sum_{x,n} {J_n\over 2}~ {\rm tr\/}
\Omega_{x,n } ~, ~~~~
n\in \pm\{ 0,\cdots,D-1\},
\end{equation}
\begin{equation}
J_{-n} \equiv J_n \equiv \lambda_{n} /\xi,~~
\lambda_\mu \equiv (\lambda,\widetilde{\lambda},
\cdots, \widetilde{\lambda}),~~
\Omega_{x,n}\equiv V_x U_{x,n} V_{x+\hat{n}}^{\dag},
\end{equation}
\begin{equation}\label{plaqsum}
S^u [U ] = S^u [\Omega ] =
\beta ~{\rm Re\/} \sum_{\Box} {\rm tr\/}\Box~.
\end{equation}
\end{mathletters}
The sum in (\ref{plaqsum}) ranges over all lattice
plaquettes ${\rm tr\/}\Box$.  In (\ref{expectdef}) quarks
are quenched by choice.  Consistency with
the Fadeev-Popov method requires $\{V_x\}$ to be
quenched---inverse partition function
$z_v^{-1}=z_v^{-1}[U]$ plays the role of
lattice Fadeev-Popov determinant.  Since as described
in Section~\ref{basics} the hopping expansion turns
$[{\Theta} ]_f $ into a composite operator of
links and transformations, in
this Section we focus on
the gaugefixed expectation value \hbox{$\langle V_y
{}_y{\rlap{$\bullet$}{\wedge}}_w V_w^{\dag}\rangle$}
of continuous
link chains \hbox{${}_y{\rlap{$\bullet$}{\wedge}}_w $}.  The
strategy is
to integrate out $\{V_x\}$ to get an expression for
\begin{equation}
{\cal V\/}_{\alpha\beta; ij}[U]
\equiv [ (V_y)_{\alpha\beta} (V_w^{\dag})_{ij}]_v.
\end{equation}
This integration is simple enough to be practical because
gauge symmetries relate
\hbox{${\cal V\/}$} to continuous link chains and
$z_v^{-1}$ suppresses all disconnected link chain loops.  Then
gaugefixing can be viewed as an operator
insertion of \hbox{${\cal V\/}$} into
\begin{equation}\label{totdef}
\langle V_y ({}_y{\rlap{$\bullet$}{\wedge}}_w ) V_{w}^{\dag}
\rangle_{\alpha j}= \Bigl[
[ V_y {}_y{\rlap{$\bullet$}{\wedge}}_w V_w^{\dag} ]\Bigr] =
\Bigl[
({}_y{\rlap{$\bullet$}{\wedge}}_w)_{\beta i} ~
{\cal V\/}_{\alpha\beta;ij}[U]
\Bigr]_u ~,
\end{equation}
the usual gauge invariant lattice expectation value.

	We will show that in the $\xi\propto 1/N$ and
$N\to\infty$ limit
\begin{equation}\label{geoexpand2}
[(V_y)_{\alpha\beta} (V^{\dag}_w)_{ij} ]_v
= \delta_{\alpha j}
\sum_{{}_w{\widetilde{\rlap{$\bullet$}{\wedge}}}_y}
A_{{}_w{\widetilde{\rlap{$\bullet$}{\wedge}}}_y}~
({}_w{\widetilde{\rlap{$\bullet$}{\wedge}}}_y
)_{i\beta } ~~~~~(N\to\infty ),
\end{equation}
where the sum ranges over a complete set of {\it selfavoiding\/}
continuous link chains
${}_w{\widetilde{\rlap{$\bullet$}{\wedge}}}_y$.  Orthonormality
of the ${}_w{\widetilde{\rlap{$\bullet$}{\wedge}}}_y$
with respect to inner product
\begin{equation}\label{measureu}
\langle \Theta_1 \vert \Theta_2 \rangle \equiv
\int [dU]~{1\over N}~{\rm tr\/}\bigl(\Theta_1^{\dag}
\Theta_2\bigr)~, ~~~~
[dU]\equiv \prod_{x,\mu} dU_{x,\mu}
\end{equation}
identifies coefficients
$A_{{}_w{\widetilde{\rlap{$\bullet$}{\wedge}}}_y}$
with $\beta =0$ $\lambda$-gauge expectation values:
\begin{equation}
A_{{}_w{\widetilde{\rlap{$\bullet$}{\wedge}}}_y} =
{1\over N^2} \int [dU] ~{\rm tr\/}
\bigl( {}_y{\widetilde{\rlap{$\bullet$}{\wedge}}}_w
[V_w^{\dag} V_y ]_v\bigr)
= {1\over N^2}{\rm tr\/}\langle
V_y {}_y{\widetilde{\rlap{$\bullet$}{\wedge}}}_w
V_w^{\dag}\rangle_{\vert \beta =0}.
\label{geoexpand3}
\end{equation}

 	If $y=w=x$,
the only continuous selfavoiding chain
is \hbox{${}_x{\widetilde{
\rlap{$\bullet$}{\wedge}}}_x ={1}$}.  Since
\hbox{$V_x V^{\dag}_x={1}$}, Eq.~(\ref{geoexpand3}) implies
\hbox{$A_{{}_x{\widetilde{\rlap{$\bullet$}{\wedge}}}_x}
=1/N $}.  Hence \hbox{$ [
(V_x)_{\alpha\beta} (V^{\dag}_x)_{ij}]_v$} is independent
of $[U]$ and
\begin{equation}\label{eq:norenorm}
\langle(V_x)_{\alpha\beta} (V^{\dag}_x)_{ij}\rangle =
[ (V_x)_{\alpha\beta} (V^{\dag}_x)_{ij} ]_v =
{1\over N}~\delta_{\alpha j}\delta_{\beta i} =
\int dV~V_{\alpha\beta} V^{\dag}_{ij}~.
\end{equation}
Eq.~(\ref{eq:norenorm}) implies gaugefixing does not affect
closed link loops $\bigcirc_x$ since
\begin{equation}
\langle\bigl(V_x\bigcirc_x V^{\dag}_x\bigr)_{\alpha j}\rangle
= {1\over N}~\delta_{\alpha j}~[{\rm tr\/}\bigcirc_x]_u
= \langle (\bigcirc_x)_{\alpha j}\rangle~.
\label{circfac2}
\end{equation}

\subsection{Link Operator Definitions}
\label{sec:defs}

	A continuous link chain or randomwalk path
may {\it selfintersect\/}(touch itself at right angles),
{\it recur\/}(touch itself at $0^o$ or $180^o$),
or {\it backtrack\/}(immediate recurrence as in the
sequence ``$\cdots U_{x,\mu}U_{x+\hat{\mu},-\mu}\cdots$'').  A
{\it selfavoiding\/} link chain or randomwalk path
does not selfintersect or recur.

	A tree $T$ is a continuous randomwalk path
which is nonbacktracking but may be
selfintersecting and otherwise recurrent.  A branch $B$ is
a continuous, possibly selfintersecting and definitely recurring
randomwalk loop enclosing a zero area minimal surface.  A
tip of $B$ is a backtracking
subsegment of $B$.  Every $B$ has at least
one tip, perhaps more.  Every randomwalk path is
a sequence of trees and branches.

	Continuous link chains
extending from $y$ to $w$, not necessarily straight or selfavoiding,
are denoted by
\begin{equation}\label{def:ywline}
{}_y{\rlap{$\bullet$}{\wedge}}_w \equiv U_{y,\mu}U_{y+\hat{\mu},\nu}\cdots
U_{w-\hat{\delta},
\delta}~, ~~~~_y{\rlap{$\bullet$}{\wedge}}_w^{\dag}
= {}_w{\rlap{$\bullet$}{\wedge}}_y
\end{equation}
where ``$\cdots$'' denotes a continuous but not necessarily
selfavoiding link chain.  Selfavoiding
link chains
\hbox{${}_y{\widetilde{\rlap{$\bullet$}{\wedge}}}_w$} are accented
with ``$\widetilde{~}$.''
Examples of recurrent and selfintersecting link chains are
\hbox{$\underline{\Box}$}, a continuous chain which traces out
a unit box and recurs
at one side, and \rlap{${}^{\diamondsuit}$}{${}_\wedge$}, which
wraps around a unit box and selfintersects at one corner.  Link
chain loops $_x{\rlap{$\bullet$}{\wedge}}_x$ are also
denoted by $\bigcirc$.  The unit square chain is $\bigcirc \equiv\Box$,
whose
trace is the plaquette.  If $\bigcirc$ begins and ends at $x$, then
it is denoted $\bigcirc_x$.

	If the path traced out by $\bigcirc_x$ is a branch, then
it is denoted $\bigcirc_x^B$.  Since
\hbox{$U_{x,\mu}U_{x+\hat{\mu},-\mu} = {1}$}
\begin{equation}\label{unityb}
\bigcirc_x^B~=~{1}~~~~~\forall
\bigcirc_x^B~.
\end{equation}
Factorizing link chains
into a product of tree and branch subsegments yields
\begin{equation}
{}_y{\rlap{$\bullet$}{\wedge}}_w= \bigcirc_y^B
{}_y{\rlap{$\bullet$}{\wedge}}_x^T \bigcirc_x^B \cdots
\bigcirc_{z}^B {}_z{\rlap{$\bullet$}{\wedge}}_{w}^T \bigcirc_{w}^B
= {}_y{\rlap{$\bullet$}{\wedge}}_w^T ~.  \label{tbdecomp}
\end{equation}
Hence we will assume that
\hbox{${}_y{\rlap{$\bullet$}{\wedge}}_w
= {}_y{\rlap{$\bullet$}{\wedge}}_w^T$} unless
specifically noted.

	By $SU(N)$ identities such as Schur's lemma,
\begin{equation}\label{schur}
\int dU~ D^{(\overline{\nu})}_{ij}(U)~D^{(\nu')}_{\alpha\beta}(U ) =
{1\over {\rm dim\/}(\nu)} \delta^{\nu,\nu'}~
\delta_{i\beta}\delta_{j\alpha}~,
\end{equation}
continuous link chains are orthonormal with respect to
(\ref{measureu}).  However some
disconnected chains mix with each other.  For example, a
{\it non\/}orthonormal
basis for gauge invariant link operators is
\begin{equation}\label{basis0}
{\cal B\/}_o \equiv
\bigl\{ 1 , \{{\rm tr\/}\bigcirc\} ,
\{{\rm tr\/}\bigcirc {\rm tr\/}{\bigcirc'}\} ,
\cdots \bigr\}~{1}~, ~~~{\rm tr\/}{1} = N
\end{equation}
where $\{\bigcirc\}$ is the set of all
nonbacktracking link loops in the lattice.  It is possible
to orthonormalize ${\cal B\/}_o$
by constructing
irreducible combinations out of its link operators which mix.

\subsection{Residual Gauge Symmetries and the Link Expansion}
\label{linkexpansion}

	QCD gauge transformations are
\begin{equation}\label{qcdgauge}
U_{x,\mu} \to Q_x U_{x,\mu } Q_{x+\hat{\mu}}^{\dag}~,
{}~~ V_x \to V_x~, ~~\psi_x\to Q_x\psi_x~.
\end{equation}
While $S^v$ breaks QCD gauge
symmetry, $S^u$, $S^f$, and $S^v$ are invariant under
\begin{equation}\label{higgsgauge}
U_{x,\mu} \to R_x U_{x,\mu } R_{x+\hat{\mu}}^{\dag},
{}~~ V_x \to V_x R^{\dag}_x, ~\psi_x\to
R_x\psi_x~ .
\end{equation}
We will refer to~(\ref{qcdgauge}) as ``$Q$'' transformations and
(\ref{higgsgauge}) as ``$R$'' transformations.  $R$ is a symmetry
of \hbox{$[\Theta]_{\theta\vert \theta\in
\{f,v,u\}}$} (and hence $\langle\Theta\rangle$) whereas
$Q$ is a symmetry
of only \hbox{$[\Theta]_{\theta\vert\theta\in \{f,u\}}$}.  By
Elitzur's theorem~\cite{elitzer} any $R$-variant
operator such as $V_x$ and
$U_{x,\mu }$ has zero $\langle\Theta\rangle$
expectation value.  $R$-invariant operators,
including $Q$-{\it variant\/} ones like
$\Omega_{x,\mu}$, are not suppressed.

	Global transformation ``$L_{XY}$'' where
\begin{equation}\label{globalsym}
U_{x,\mu} \to U_{x,\mu }~,
{}~~ V_x \to L_{XY} V_x~, ~~ \psi_x\to \psi_x
\end{equation}
is a symmetry of
\hbox{$[\Theta]_{\theta\vert\theta\in \{f,v,u\}}$}.  $L_{XY}$
is equivalent to
global color symmetry.  The
$XY$ designation is because in the
$\beta\to\infty$ limit the system approaches the
$XY$ model, which has $L_{XY}$ symmetry.

	Under \hbox{$U_{x,\mu} \to
U_{x,\mu}' = Q_x U_{x,\mu} Q_{x+\hat{\mu}}^{\dag}$} partition
function $z_v[U]$ obeys
\begin{equation}
z_v [U'] = \int [dV]  {\rm e\/}^{-S^v[\Omega'] } =
\int [d(VQ^{\dag})]  {\rm e\/}^{-S^v[\Omega ]} = z_v [U].
\label{zgaugeinv}
\end{equation}
As a $Q$ invariant functional of
links $z_v$, the inverse of the Fadeev-Popov determinant,
is expandable in terms of
gauge invariant link structures.  Such structures are
comprised of links joined together into closed networks
with gauge invariant bonds $\delta_{ij}$
and \hbox{$\epsilon_{i_1\cdots i_N}$}.  Let us
take $N\mapsto\infty$ or $U(1)$ and throw
out the latter.  Then following (\ref{basis0}), with
\hbox{$\bigcirc\ne 1$} and \hbox{$\bigcirc '\ne 1$},
\begin{mathletters}
\begin{equation}\label{expandzv}
z_v[U]= \bigl( 1 +
\sum_{\bigcirc} Z_v^\bigcirc {\rm tr\/}\bigcirc
+\sum_{\bigcirc'\ne \bigcirc^{\dag}}
Z_v^{\bigcirc\bigcirc'} {\rm tr\/}
\bigcirc {\rm tr\/}\bigcirc' +\cdots \bigr)
Z^o_v ,
\end{equation}
\begin{equation}
Z^o_v \equiv
\int [dU] z_v[U] =  Z^{1}_v  +
\sum_{\bigcirc\ne 1}
Z_v^{\bigcirc\bigcirc^{\dag}}{\rm tr\/}
\bigcirc{\rm tr\/}\bigcirc^{\dag} +\cdots
\end{equation}
\end{mathletters}
The reason for factorizing out $Z^o_v$ in (\ref{expandzv}) and
the reason it can be expanded in terms of
zero area (colorless) link structures is because
$Z_v^o$ is the colorless part of $z_v[U]$.  Since it is colorless,
$Z_v^o$ survives at $\beta=0$.

	If $W$ is in the fundamental representation and
$\chi^{(\nu )}$ the character of irreducible representation
$\nu$, the character
expansion of ${\rm e\/}^{z {\rm Re\/}{\rm tr\/} W}$ is given by
\begin{mathletters}
\begin{equation}
b_{\overline{\nu}}(z) = b_{\nu}(z) \equiv
\int dW~\chi^{(\nu )}(W )~
{\rm e\/}^{{z\over 2} {\rm tr\/}(W+W^{\dag}) },
\end{equation}
\begin{equation}
{\rm e\/}^{{z\over 2} {\rm tr\/}(W+W^{\dag}) }
= \sum_{\nu } b_\nu (z ) ~\chi^{(\nu )}(W ) ,
{}~~~ f_\nu \equiv {b_\nu\over N b_1}.
\end{equation}
Since by (\ref{svdef}) \hbox{${\rm e\/}^{-S^v[\Omega ]}
=\prod_{x,\mu}
{\rm e\/}^{J_\mu {\rm Re\/}{\rm tr\/}\Omega_{x,\mu}}$},
\begin{equation}\label{zcharacter}
{\rm e\/}^{-S^v[\Omega]} = { Z^o_v }
\sum_{\{\nu_{x,\mu}\}}
\prod_{x,\mu } N f_{\nu_{x,\mu}} (J_\mu) \chi^{(\nu_{x,\mu})}
(\Omega_{x,\mu} )
\end{equation}
where the reason for factoring out $Z^o_v$ will be
apparent shortly and where
\begin{equation}\label{eq:fx}
/\Theta /~ \equiv
{\int dW~ \Theta ~
{\rm e\/}^{{1\over 2} z({\rm tr\/}(W + W^{\dag})} \over
\int dW'~  {\rm e\/}^{{1\over 2}
z({\rm tr\/}(W' + W'^{\dag})} }~, ~~~
f_N( z) \delta_{ij}~\equiv~ /W_{ij}/~,
\end{equation}
\begin{equation}\label{solidbound}
\Vert f_N(z)\Vert \le 1, ~~~
\lim_{z\to 0}  f_N(z) = {z\over 2N}, ~~~
f_N (\infty)= 1 ~~~~({\rm finite\/}~N).
\end{equation}
\end{mathletters}
The first inequality in (\ref{solidbound}) follows from
$\Vert {\rm tr\/} W\Vert \le N$;
the second from expanding
${\rm e\/}^{z{\rm Re\/}{\rm tr\/} W}$ in $z$;
the third since the integrands are dominated
by Max${\rm Re\/}{\rm tr\/} W$ when $z\to\infty$.  Integrating
both sides of (\ref{expandzv})
over $[dU]$, interchanging integration order
\hbox{$[dU][dV]\to [dV][dU]$} and
changing variables $[dU]$ to $[d\Omega ]$ yields
\begin{equation}
Z^o_v  = \int [dU] [dV]~
\prod_{x,\mu} {\rm e\/}^{J_\mu {\rm Re\/}
{\rm tr\/} \Omega_{x,\mu} }
= \prod_{x,\mu} b_1 (J_\mu )~. \label{z1answ}
\end{equation}

	 $L_{XY}$ symmetry implies
\begin{equation}\label{geoexpand1}
\bigl[(V_y)_{\alpha\beta} (V^{\dag}_w)_{ij}\bigr]_v =
\delta_{\alpha j} ~ t_{i\beta } [U;w,y]~.
\end{equation}
Since $t[U;w,y]$ transforms under local $R$
like \hbox{$
[(V_y)_{\alpha\beta} (V^{\dag}_w)_{ij} ]_v$},
at $y$ it transforms like $N$, at $w$ like $\overline{N}$,
and the links away from $y$ and $w$ must be bonded gauge
invariantly either by the $SU(N)$ identity tensor or the
completely antisymmetric $\epsilon$ tensor.

	To appreciate that (\ref{geoexpand2}) is not a
selfevident result---it's valid only in the $N\to\infty$ limit and
orthogonal trace approximation to be described---consider
the $U(1)$ case where
\begin{mathletters}
\begin{equation}
U_{x,\mu} \equiv {\rm e\/}^{i\phi_{x,\mu}},~~
V_x\equiv {\rm e\/}^{i\theta_x},~~
S^v =-\sum_{x,\mu } {J}^\mu
\cos\bigl(\theta_x - \theta_{x+\mu} +\phi_{x,\mu}\bigr).
\end{equation}
The $U(1)$ character expansion is given by
\begin{equation}\label{cexpand}
{\rm e\/}^{z\cos\theta } = I_0 (z) \sum_{n=-\infty}^\infty
t_n (z) {\rm e\/}^{-in\theta },
{}~~~ \chi^{(n)}({\rm e\/}^{-i\theta}) = {\rm e\/}^{-in\theta},
\end{equation}
\begin{equation}\label{characteru1}
I_{-n}(z) =
I_n (z) = {1\over 2\pi}
\int_{-\pi}^{\pi} d\theta~{\rm e\/}^{in\theta
+z\cos(\theta )} , ~~~
t_n \equiv {I_n/ I_0},
\end{equation}
\begin{equation}\label{firstu1}
\overline{z}_v [\phi] = z_v/Z^o_v = \sum_{ \{n_{x,\mu}\}'}
\prod_{x,\mu} t_{n_{x, \mu}}
(J_\mu)~{\rm e\/}^{-in_{x,\mu}\phi_{x,\mu}}~,
\end{equation}
\begin{equation}
\{n_{x,\mu}\}'\equiv
\bigl\{ \{n_{x,\mu }\}\vert
\sum_{m} n_{x,m} = 0\bigr\}~,
\end{equation}
\begin{equation}\label{charactertiles}
[V_y V_w^{\dag}]_v = \overline{z}_v^{-1} [\phi]
\sum_{ \{n_{x,\mu}\}'' }
\prod_{x,\mu} t_{n_{x, \mu}} (J_\mu)~
{\rm e\/}^{-in_{x,\mu}\phi_{x,\mu}}~,
\end{equation}
\begin{equation}\label{contcond}
\{n_{x,\mu}\}'' \equiv
\bigl\{ \{n_{x,\mu }\}\vert
\sum_{m} n_{x,m} =\delta_{x,y}-\delta_{x,w} \bigr\}~.
\end{equation}
\end{mathletters}
The $U(1)$ identity used to obtain (\ref{firstu1}-\ref{contcond}) is
\begin{equation}\label{u1property}
\int dV_x ~\prod_m \chi^{(\nu_{x,m})} (\Omega_{x,m})
= \delta\bigl(\sum_{m'} \nu_{x,m'} \bigr)~\prod_{m}
\chi^{(\nu_{x,m})}(U_{x,m }V_{x+\hat{m}}^{\dag})
\end{equation}
where
\hbox{$\chi^{(\nu_{x,-\mu})}(\Omega_{x,\mu})\equiv
\chi^{(\nu_{x-\hat{\mu},-\mu})}(\Omega_{x-\hat{\mu},\mu}^{\dag} )$}.

	Because $U(1)$ character coefficients do not satisfy
\begin{equation}\label{fact0}
t_n (z) = \bigl(t_1 (z)\bigr)^n ~,
\end{equation}
link loops in the
numerator of $[V_y V_w^{\dag}]_v$ which touch open chains
\hbox{${}_w{\widetilde{\rlap{$\bullet$}{\wedge}}}_y$} are
not cancelled by corresponding $z_v$
loops and (\ref{geoexpand2}) does not
apply to $U(1)$.  Formula~(\ref{geoexpand2})
only applies to systems
consistent with both (\ref{u1property})
and (\ref{fact0}).

	The $SU(N)$ version of (\ref{u1property}) is
complicated because there is more than one way
of making a singlet out of
\hbox{$\nu_{-D+1}\otimes \cdots\otimes \nu_{D-1}$.}  $SU(\infty)$
traces, on
the other hand, obey an analog of (\ref{u1property}):
\begin{mathletters}
\begin{eqnarray}\label{sitetraceid}
&&\int dV_x \prod_{m=-D+1}^{D-1}
{\rm tr\/}^{n_{x,m}} (V_x O_{x,m}) ~{\rm tr\/}^{l_{x,m}} (O_{x,m}^{\dag}
V_{x}^{\dag}) \\
 && ~~~~~~~~ =
\delta(\overline{n}_{x,D-1}-\overline{l}_{x,D-1})
\sum_{\sigma_x\in S_{\overline{n}_{x,D-1}} }
\prod_{j_x=0}^{\overline{n}_{x,D-1}} \Biggl[{{\rm tr\/}( O_{x,M_x(j)}
O^{\dag}_{x,M_x( \sigma (j))} ) \over N}\Biggr] \nonumber
\end{eqnarray}
where
\begin{equation}
O_{x,m} = \cases{ U_{x,\mu} V_{x+\hat{\mu}}^{\dag} & $m=\mu \ge 0$;\cr
U_{x-\hat{\nu},\nu}^{\dag} V_{x-\hat{\nu}}^{\dag} & $m=-\nu \le 0$;\cr}
\end{equation}
\begin{equation}
n_{x,m} =\cases{ n_{x,\mu} & $m=\mu $;\cr
l_{x-\hat{\nu},\nu} & $m=-\nu $;\cr} ~~~~~
l_{x,m} =\cases{ l_{x,\mu} & $m=\mu $;\cr
n_{x-\hat{\nu},\nu} & $m=-\nu $;\cr}
\end{equation}
\begin{equation}
\overline{n}_{x,m} \equiv \sum_{m'=-D+1}^m n_{x,m'}, ~
M_x = \cases{1-D &$0\le j\le n_{x,1-D}$;\cr
{}~~ ~~\vdots &~~ ~~ $\vdots$\cr
D-1 &$\overline{n}_{x,D-2} < j\le \overline{n}_{x,D-1}$.\cr}
\label{combinatoricint}
\end{equation}
\end{mathletters}
\noindent{}$S_p$ is the permutation group on $p$
elements.  Eq.~(\ref{sitetraceid}) says that the numerators
and denominators of $t[U;w,y]$ are comprised of chains
made by joining links
from the trace expansion of ${\rm e\/}^{-S^v}$ in
all possible permutations consistent with
\hbox{$\{n_{x,m},l_{x,m}\}$} conservation at each site.  In
Section~\ref{trexpand} we will find that relevant
coefficients of the link expansion
satisfy (\ref{fact0}) and that the $SU(\infty)$ analog
of (\ref{firstu1}-\ref{contcond})
is consistent with Eq.~(\ref{geoexpand2}).

\subsection{Trace Expansion and Orthogonality Approximation}
\label{trexpand}

	We will call the use of
(\ref{sitetraceid}) the ``trace orthogonality
approximation.''  The reason its use is an approximation even in
the $N\to\infty$ limit is as follows.  If $2 \le N <\infty$,
an exact group integral formula is
\begin{eqnarray}
\int dV && V_{i_1 j_1} V^{\dag}_{\alpha_1\beta_1}
V_{i_2 j_2} V^{\dag}_{\alpha_2\beta_2} =
{1\over N^2 -1} \Bigl[ \nonumber \\
 && ~~\bigl(\delta_{i_1\beta_1} \delta_{j_1 \alpha_1}
\delta_{i_2\beta_2} \delta_{j_2 \alpha_2} - {1\over N}
\delta_{i_1\beta_1} \delta_{j_1 \alpha_2}
\delta_{i_2\beta_2} \delta_{j_2 \alpha_1}\bigr)
+ \Bigl( {\alpha_1\leftrightarrow\alpha_2
\atop \beta_1\leftrightarrow\beta_2 }\Bigr) \Bigr].\label{sampleid}
\end{eqnarray}
Since the second and fourth terms in the RHS of (\ref{sampleid})
are suppressed by $1/N$ relative to the other terms, (\ref{sampleid})
agrees with (\ref{sitetraceid})
in the $N\to\infty$ limit.  However, using (\ref{sampleid})
in place of (\ref{sitetraceid}) to compute
the leading ${\cal O\/}(J_\mu)$ hopping expansion term of
$A_{\underline{\Box}}$, the
Eq.~(\ref{geoexpand2}) coefficient where
$\underline{\Box}$ is a continuous five-link chain
which traces out
a unit square and recurs at one side, yields
\begin{equation}\label{nonvanishingrecurrence}
A_{\underline{\Box}} \approx
{1\over N} \Bigl({J_\mu\over 2N}\Bigr)^5
\Bigl[ 1 - 1 - 1\Bigr]
= - {1\over N} \Big({J_\mu\over 2N}\Bigr)^5 \ne 0.
\end{equation}
The first two terms of the middle expression come
from the numerator of $t[U;w,y]$; the
third from a denominator plaquette and a numerator link.  Since
$J\propto N$ (as described below),
\hbox{$A_{\underline{\Box}} = {\cal O\/}(1/N)$} if we
take $N\to\infty$
{\it after\/} resummation of the trace expansion.  On the other
hand, $A_{\underline{\Box}}
=0$ if one uses
Eq.~(\ref{sitetraceid}), valid
to leading ${\cal O\/}(1/N)$, to integrate out
the $[dV]$ (as described below).  The latter is tantamount to
taking $N\to\infty$ {\it before\/} resummation.  The discrepancy
arises because the leading
\hbox{${\cal O\/}(1/N)$} contribution to $A_{\underline{\Box}}$ comes
from {\it sub\/}leading terms of (\ref{sitetraceid}) which have been
thrown out.  We have not been able to improve
the trace orthogonality approximation to account properly
for coefficients of recurrent chains.

	The orthogonal trace approximation leads to the
following for
\begin{mathletters}
\begin{equation}\label{generator}
Z[L,K] = \int dW~{\rm e\/}^{{1\over 2}{\rm tr\/}[ L
W+ W^{\dag} K]}, ~~~~
Z[J,J] = b_1 (J).
\end{equation}
Expanding in $L$ and $K$,
taking $N\to\infty$, imposing
\begin{equation}\label{badidentity}
\int dW~ {\rm tr\/}^n (LW) {\rm tr\/}^l (W^{\dag} K) =
n!~  \Bigl[{{\rm tr\/}(LK)\over
N}\Bigr]^n~ \delta_{nl} ~~~ (N\to\infty)
\end{equation}
and resumming gives
\begin{equation}
\widetilde{Z}[L,K] = \sum_{l=0}^\infty \sum_{n=0}^l
2^{-l} \int dW~ {{\rm tr\/}^n (LW)
{\rm tr\/}^{l-n}(W^{\dag} K)\over n! ~ (l-n)!}
= {\rm e\/}^{{\rm tr\/}(L K)\over 4N}. \label{approxform}
\end{equation}
\end{mathletters}
\noindent{}(\ref{badidentity}) is a special case of
(\ref{sitetraceid}).  We change notation
from $Z\mapsto \widetilde{Z}$ to
emphasize that the interchange of integration and
resummation leads to a discrepancy,
described below,
between $\widetilde{Z}$ and the original $Z$.

	Using $\widetilde{Z}$ as a generating function
yields (``$/~/$'' is defined in (\ref{eq:fx}).)
\begin{mathletters}
\begin{equation}
F^{(l,n)}_{i_1 j_1;
\cdots; i_l j_l \atop k_1 o_1;\cdots; k_n o_n}
\equiv ~/ W_{i_1 j_1}^{\dag}\cdots W_{i_l j_l}^{\dag}
W_{k_1 o_1}\cdots W_{k_n o_n}/~,
\end{equation}
\begin{equation}\label{onef}
2 {\partial\log \widetilde{Z}
\over \partial L_{j i}}_{\vert {L= J
\atop K = J}} =
2 {\partial\log \widetilde{Z}
\over \partial K_{j i}}_{\vert {L= J
\atop K = J}} =
F^{(0,1)}_{ij} = F^{(1,0)}_{ij} = \delta_{ij}~f_N(J),
\end{equation}
\begin{equation}\label{badf}
f_N(J) = {J\over 2N}~~~~~~~ (J<< 2N, ~N\to\infty).
\end{equation}
Since following (\ref{approxform})
\begin{equation}\label{factapprox}
{\partial^{n+l}\log \widetilde{Z}\over \partial K_{j_1 i_1}\cdots
\partial K_{j_l i_l}\partial L_{o_1 k_1}\cdots\partial L_{o_n k_n}
} = 0 ~~~~({\rm if\/}~ n ~{\rm or\/}~ l>1),
\end{equation}
\begin{equation}\label{generateit}
F^{(l,n)}_{i_1 j_1; \cdots; i_l j_l \atop k_1 o_1;\cdots; k_n o_n}
(J) - \prod_{p=1}^{n+l} \delta_{i_p j_p} f_N(J) ~=~ 0
\end{equation}
by induction.  Combining (\ref{onef}) and (\ref{generateit}) yields
\begin{equation}\label{eq:factorize}
{1\over N^{n+l}} /
{\rm tr\/}^n W ~{\rm tr\/}^l W^{\dag} / = f_N^{n+l} (J)~
{}~~~~ (n\ge 0,~l\ge 0)~.
\end{equation}
\end{mathletters}
Eq.~(\ref{eq:factorize}) is our preliminary $SU(\infty)$
analog of (\ref{fact0}) which we will now improve.

	Formula~(\ref{badf}) violates the inequality
of (\ref{solidbound}) if $J > 2N$ because
$\widetilde{Z}$ is {\it not\/} valid to
leading ${\cal O\/}(N)$.  Rather,
following (\ref{badidentity}), it is only asymptotically
valid as ${N\to\infty}$.
Since
(\ref{approxform}) is analytic in ${\rm tr\/}(LK)$, the violation
implies $\widetilde{Z}\ne Z$.   This
discrepancy arises from
applying (\ref{badidentity}) to (\ref{approxform})
on $n\to\infty$ (including $n\ge N$) terms
before resummation.

	An exactly solvable limit is $N\to\infty$ and
$J\propto N$.  In this double limit,
achieved by identifying
the gauge parameter $\xi$ with $\xi\propto 1/N$ so that
\begin{equation}\label{eq:limits}
N\to\infty,~~~\xi\to 0,~~~ {J_\mu \over 2N} \equiv \lambda_\mu~,
\end{equation}
the $N$ dependence cancels out leaving a
nontrivial $\lambda$-dependent
solution.  The exact generating function is~\cite{witten}
\begin{mathletters}
\begin{equation}\label{newzdef}
\lim_{N\to\infty}
Z[2N\ell, 2N\kappa ]  \equiv
\int dW~{\rm e\/}^{N{\rm tr\/}(\ell W + \kappa W^{\dag})}
\equiv  {\rm e\/}^{N^2 w(\ell,\kappa )},
\end{equation}
\begin{equation}
w(\ell,\kappa ) = \cases{ ~~~~~ \ell\kappa &
$\sqrt{\ell\kappa}\le {1\over 2}$;\cr
2\sqrt{\ell\kappa} -{3\over 4} - {1\over 4}\log (4\ell\kappa )&
$\sqrt{\ell\kappa} > {1\over 2}$;\cr}
\end{equation}
\begin{equation}\label{fnformula}
\lim_{N\to\infty\atop \xi\to 0} f_N (J)  =
{\partial w\over \partial \ell }_{\ell =\kappa =\lambda } =
\cases{ ~~~ \lambda
&$\lambda \le {1\over 2} $;\cr 1-{1\over 4\lambda }
&$\lambda \ge {1\over 2}$;\cr}
\end{equation}
\begin{equation}\label{factorizer0}
{1\over N^2}~/{\rm tr\/}^2 W / - f_N(J) =
{1\over N^2} {\partial^2 w\over\partial \alpha^2 } =
{\cal O\/}\bigl(
{1\over N^2}\bigr)~.
\end{equation}
By induction
\begin{equation}~\label{factorizer}
{1\over N^{n+l}} /
{\rm tr\/}^n W~ {\rm tr\/} ^l W^{\dag} / -  f_N^{n+l}(J)~=~
{\cal O\/}\bigl( {1\over N^2}\bigr).
\end{equation}
\end{mathletters}
Since the LHS terms in (\ref{factorizer0}) and
(\ref{factorizer}) are ${\cal O\/}(1)$,
property~(\ref{eq:factorize}) is valid when $J\equiv 2N\lambda$
in the $N\to\infty$ limit.

	The link expansion of $t[U;w,y]$ follows
from trace expansion
\begin{equation}\label{dexpand}
{\rm e\/}^{z {\rm Re\/}{\rm tr\/}\Omega } =
\sum_{n=0}^\infty\sum_{l=0}^\infty
\tau (n,l;z) {{\rm tr\/}^n\Omega\over n!}
{{\rm tr\/}^l\Omega^{\dag} \over l!}  ~.
\end{equation}
Taylor expansion of
${\rm e\/}^{z{\rm Re\/}{\rm tr\/}\Omega}$, which
defines the exponential operator, implies
$\tau (n,l;J) = (J/2)^{n+l}$.  In fact, this
identification is only asymptotically consistent with
using (\ref{sitetraceid}) in the limit
$J << 2N $ and $N\to\infty$.  As $J\propto N\to\infty$,
integral formula (\ref{sitetraceid})
cannot be interchanged with
resummation.  If we insist on applying (\ref{sitetraceid})
before resumming the trace expansion, $\tau (n,l;J)$ must be
redefined so that
correlation functions evaluated using (\ref{sitetraceid})
and (\ref{dexpand}) are consistent
with correlation functions evaluated directly from
$Z$ of Eq.~(\ref{newzdef}).  The matching condition
\begin{equation}
\sum_{n=0}^\infty {\tau(n,n+k;J )\over n!} =
b_1 (J)~ N^k ~ f_N^k (J ) =
\sum_{n=0}^\infty {\tau(n+k,n;J )\over n!}
\end{equation}
implies when $N\to\infty$
\begin{equation}\label{tauvalues}
\tau(n,l;J ) =
N^{l+n}~ f_N^{l+n}(J) \cases{ ~~~~~~1 &$J<< 2N$;\cr
{\rm e\/}^{N^2 [w(\lambda ,\lambda )-\lambda^2 ]}&$J=2N\lambda$.\cr}
\end{equation}
The $SU(\infty)$ analog of (\ref{fact0}),
\begin{equation}\label{eq:factorid}
\tau(n,l+k; J) =
\tau(n+k,l; J) = N^k~ f_N^k(J)~ \tau (n,l;J) ~,
\end{equation}
will be referred to as ``factorization.''

	Applying Eqs.~(\ref{sitetraceid}) and (\ref{dexpand}) leads
to the $SU(\infty)$ version
of (\ref{firstu1}-\ref{contcond}).  The $[dU]$ integral of
Eq.~(\ref{geoexpand3}) for
$A_{{}_w{\widetilde{\rlap{$\bullet$}{\wedge}}}_y}$ projects the
the numerator and denominator of $t[U;w,y]$ down to
the colorless part
except along \hbox{${}_w{\rlap{$\bullet$}{\wedge}}_y^{\dag}$}.  At
other sites the denominator contributions
cancel the corresponding numerator contributions.  Each link
of \hbox{${}_w{\rlap{$\bullet$}{\wedge}}_y$} coming from
a (\ref{sitetraceid}) integration costs a factor of
$1/N$.  Additionally there
is an overall factor
of \hbox{${\rm tr\/}
( {}_w{\widetilde{\rlap{$\bullet$}{\wedge}}}_y^{\dag}
{}_w{\widetilde{\rlap{$\bullet$}{\wedge}}}_y)/N^2 = 1/N$}.  Hence
\begin{mathletters}
\begin{equation}\label{ntoinftysoln}
A_{{}_w{\widetilde{\rlap{$\bullet$}{\wedge}}}_y} =
{\wedge_{\{n,x,m\}}
{\tau(n_{x,m},n_{x,m}+p_{x,m},J_m )\over n_{x,m}! ~ N^{p_{x,m}} }
\over
N \wedge_{\{l,y,o\}}
{\tau(l_{y,o},l_{y,o},J_o )\over l_{y,o}! } } =
{1\over N}
\prod_{\{x,m\}\in {}_w{
\widetilde{\rlap{$\bullet$}{\wedge}}}_y} f_N (J_m )
\label{selfavoidinggraph}
\end{equation}
where
\begin{equation}
\wedge_{\{n,x,m\}}\equiv \sum_{\{n_{x,m}\}}\prod_{\{x,m\}}~,~~~~
p_{x,m} = \cases{1&if $\{x,m\}\in
{}_w{\widetilde{\rlap{$\bullet$}{\wedge}}}_y$;\cr
0&if $\{x,m\}\notin {}_w{
\widetilde{\rlap{$\bullet$}{\wedge}}}_y$.\cr}
\end{equation}
\end{mathletters}

	Let $a( \Theta )$ be the coefficient of link chain
$\Theta$ in the numerator or denominator of $t[U;w,y]$.  Consider
again $A_{\underline{\Box}}$ where
$\underline{\Box}$ is a link chain
which traces out
a unit square and recurs at one side.  The chain
$\underline{\Box}$ does not contribute to $A_{\underline{\Box}}$
in the same manner that
\hbox{${}_w{\widetilde{\rlap{$\bullet$}{\wedge}}}_y$} contributes to
\hbox{$A_{{}_w{\widetilde{\rlap{$\bullet$}{\wedge}}}_y}$}
because, due to factorization,
$a (\underline{\Box}) = a(-)~ a(\Box )$ so that
\begin{equation}\label{factorizeexample}
{ a (-) + a (\underline{\Box}) \over 1 + a (\Box )}
= a(-) ~ { 1 + a (\Box) \over 1 + a (\Box )}
= a (-)~.
\end{equation}
(\ref{factorizeexample}) says that
the $\underline{\Box}$ contribution
to $A_{\underline{\Box}}$
has already been counted in $A_{{\rm -\/}}$.  This implies that
\begin{equation}\label{nonselfintersectionrule}
A_{{\rm recurrent~or\/}\atop {\rm selfintersecting~chain\/}}
=~ 0~.
\end{equation}
As described in the beginning of this subsection
(\ref{nonselfintersectionrule}), the trace orthogonality
approximation result, is inconsistent with
the leading $J_\mu$ expansion contribution.  The
problem is that the leading contribution to
$A_{\underline{\Box}}$ is {\it not\/} from the leading term of
(\ref{sampleid}).  In (\ref{nonvanishingrecurrence})
the middle ``$-1$'' comes from the second
term of (\ref{sampleid})---a term which is neglected in
(\ref{sitetraceid}).  Hence it does not appear in
$A_{\underline{\Box}}$ if
(\ref{sitetraceid}) is applied to do the $[dV]$
integrals.

\subsection{Wilson Loop Picture of Gaugefixing}
\label{looppix}

	Eqs.~(\ref{totdef}), (\ref{geoexpand2}),
and (\ref{ntoinftysoln}) imply
that as $N\to\infty$ within the trace orthogonality approximation
\begin{equation}\label{looppicturep}
\langle V_y ({}_y{\rlap{$\bullet$}{\wedge}}_w )V_{w}^{\dag}
\rangle_{\alpha j}  =
\delta_{\alpha j} \sum_{{}_w{\widetilde{\rlap{$\bullet$}
{\wedge}}}_y} A_{{}_w{\widetilde{\rlap{$\bullet$}{\wedge}}}_y}(J )~
{\rm tr\/}[ {}_y{\rlap{$\bullet$}{\wedge}}_w~
{}_w{\widetilde{\rlap{$\bullet$}{\wedge}}}_y ]_u ~,
\end{equation}
\begin{equation}\label{strongformulap}
A_{{}_w{\widetilde{\rlap{$\bullet$}{\wedge}}}_y} =
{1\over N^2}{\rm tr\/}\langle
V_y {}_y{\widetilde{\rlap{$\bullet$}{\wedge}}}_w
V_w^{\dag}\rangle_{\vert \beta =0}
= \prod_{\{x,\mu\}\in {}_y{
\widetilde{\rlap{$\bullet$}{\wedge}}}_w } f_N (J_\mu )~,
\label{geoexpand3pp}
\end{equation}
where $f_N$ is given in (\ref{badf}) if $J_\mu $ is finite and
in (\ref{fnformula}) if $\xi\to 1/N$.  These formulas have the
following interpretation:
\begin{itemize}
\item By Eq.~(\ref{looppicturep}) the gaugefixed expectation
value of a link chain ${}_y{\rlap{$\bullet$}{\wedge}}_w$
is the weighted sum over all Wilson loops
made by joining to ${}_y{\rlap{$\bullet$}{\wedge}}_w$ a
selfavoiding link segment
$_w{\widetilde{\rlap{$\bullet$}{\wedge}}}_y$.  This holds at all $\beta$.
\item By (\ref{geoexpand3pp}),
weights $A_{_w{\widetilde{\rlap{$\bullet$}{\wedge}}}_y}$
are proportional to
the $\beta=0$ gaugefixed expectation value of
$_w{\widetilde{\rlap{$\bullet$}{\wedge}}}_y$.  {\it All\/}
{\it the\/} {\it  gauge\/} {\it dependence\/} {\it is\/}
{\it in \/} {\it these\/} {\it weights\/}.
\item $A_{_w{\widetilde{\rlap{$\bullet$}{\wedge}}}_y}$
is a path-product of
${\rm e\/}^{-S^v}$ trace expansion coefficients.  In the
orthogonal trace approximation (\ref{badidentity}),
nonselfavoiding operators do not contribute to
$[V^{\dag}_w V_y]_v$ and the $\beta=0$ expectation
value of nonselfavoiding links vanish.
\item Following
Eq.~(\ref{eq:factorid})
$\beta=0$ $\lambda$-gauge
expectation values satisfy factorization.  If $P_{SA}$
is a selfavoiding path
\end{itemize}
\begin{equation}\label{factorizationtheorem}
\langle \prod_{\{y,n\}\in P_{SA}} \Omega_{y,n}\rangle_{\vert
\beta=0}=
\prod_{\{y,n\}\in P_{SA}}\langle \Omega_{y,n}\rangle_{\vert
\beta=0}.
\end{equation}

	These results permit us to read off
the Wilson line $\lambda$-gauge expectation value,
which is proportional to the heavy quark
propagator~\cite{bs}.  Asymptotically it can be parametrized as
\begin{equation}
\lim_{t_E\to\infty} {1 \over N }
{\rm tr\/} \langle V_{(t_E,\vec{y})}
\widehat{U}_{\vec{y},\vec{y}+t_E\hat{0}}
V^{\dag}_{(0,\vec{y})}
\rangle =Z_H {\rm e\/}^{-V_H^{(\lambda )} \vert t_E\vert}, ~~~
x\equiv (t_E,\vec{0}).
\end{equation}
$V_H^{(\lambda )}$ can be interpreted as
\hbox{$V_H^{(\lambda )} =
M^{(\lambda )}_{q,  {\rm heavy\/}} - m_q$}.  At $\beta=0$
and $N\to\infty$,
\begin{equation}\label{sunheavyq}
Z_H = 1, ~~~ V_H^{(\lambda )} =-\log\bigl[f_N(J_0)\bigr].
\end{equation}
Hence $V_H^{(\lambda )}$ decreases with increasing $\lambda$.

\section{Hopping Expansion and Resummation}
\label{basics}

	In the $N\to\infty$ hopping
expansion of $[\Theta ]_f$, the $\psi_x$
operator hops from site to site leaving behind a trail of
\begin{equation}
(r - \rlap{$\slash$}{n} )~K_B U_{y,n}~,
{}~~~~ n\in \{\pm 0, \cdots, \pm (D-1)\}
\end{equation}
factors to mark its randomwalk path $P$.  Pauli
Exclusion is irrelevant because quark
quenching effectively requires quark paths
to selfintersect indefinitely.\footnote{The author thanks
C. Bernard for pointing this
out.}\newpage  \noindent{}Therefore the
quenched background field quark propagator is
\begin{equation}\label{qcontraction}
[\psi_x \overline{\psi}_0]_f =
\sum_{P} \prod_{\{y,n\}\in P} (r-\rlap{$\slash$}{n} )~K_B U_{y,n}
\end{equation}
where $\{y,n\}$ are the locations and directions
along $P$.  (Assume
implicit pathordering when appropriate.)  Following
Eq.~(\ref{tbdecomp}), decompose the sum over $P$ into
a sum over trees and branches.  The gaugefixed expectation
value of branches is trivial by (\ref{unityb}).  $\beta=0$
expectation values of nonselfavoiding
trees, as discussed in Section~\ref{trexpand},
are suppressed in the orthogonal trace approximation.  Thus
following (\ref{strongformulap}) Eq.~(\ref{qcontraction})
is equivalent to
\begin{equation}\label{eq:branches}
P_q\equiv \langle V_x \psi_x \overline{\psi}_0 V^{\dag}_0\rangle =
\sum_{\{T_{SA}\}} \bigl[\prod_{\{y,n\}\in T_{SA}} f_N(J_n) \bigr]
\sum_{\{B_y\vert y\in T_{SA}\}}
\prod_{\{y,n\}\in P} (r-\rlap{$\slash$}{n} ) K_B
\end{equation}
where $P$ is the selfavoiding tree $T_{SA}$ going through
sites $y$ with a branch $B_y$ at each site.

	 As there is no technique for summing selfavoiding trees
as specified in (\ref{eq:branches}), we
shall sum all trees $T$.  This
approximation is justified if $D>4$ by the fact that
the quark mass pole is determined in the $t\ge L\to\infty$
limit.  In this limit the average number of selfintersections
and recurrences~\cite{parisi}
\begin{equation}
\overline{s} ~=~
L\times ({\rm path~density\/})\sim {L^2\over (\overline{x^2})^{D/2}}
{}~=~ L^{2-D/2}
\end{equation}
vanishes if $D >4$ for randomwalk paths.
(Nonbacktracking
randomwalks
are presumably {\it at\/} {\it least\/}
as selfavoiding.)  This suggests that
the number of nonselfavoiding
paths, being negligible, may
not affect the quark mass pole.

	Whether selfavoidance affects the propagator
pole or not is analogous to whether
\begin{equation}\label{limitcond}
f(a,x)\equiv \sum_{L=0}^\infty a_L~ x^L~,~~~~
\lim_{L\to\infty} a_L = a^L
\end{equation}
guarantees $f(a,x)$ is of the form
\hbox{$f(a,x) \propto {1\over 1-a x}  +  {\rm finite\/}$}.  A
counterexample satisfying (\ref{limitcond})
is \hbox{$a_L = a^L + 1/L$} in which case
a second pole exists at $x=1$.  Hence
anomalous quark mass poles or mass shifts
stemming from the inclusion of
nonselfavoiding paths is a possibility which cannot be ruled out.

	There are three effects we wish to investigate: (i)spontaneous
chiral symmetry breaking, (ii)gauge dependence of the dispersion
relation, and (iii)proximity
of the effective quark propagator
to free particle form.  Since the source of
$(i)$ is certainly different from $(ii)$ and $(iii)$,
we may pursue them separately.  If the dispersion
relation is gauge dependent at $(r,\beta)$,
there is no reason---barring a critical point---to believe
this property would not persist to other $(r,\beta )$ values
since $(r,\beta)$ are unrelated to
gaugefixing.  To study (ii) and (iii) we
set $(r=1,\beta =0)$ where
we can solve for the full quark propagator and its dispersion
relation.  In this limit, spontaneous symmetry
breaking is absent.  To study the combination
of spontaneous symmetry breaking and gauge dependence, we set
$(r=0,\beta =0)$ where
we can solve for $M_q^{(\lambda )}$.

\subsection{$N\to\infty$, $r=1$, $\beta = 0$ Quark Propagator}
\label{mcpsibarpsi}

	Since \hbox{$(1-\gamma_\mu ) (1+\gamma_\mu ) = 0$}, $r=1$
quarks cannot backtrack and the hopping expansion cannot
generate any zero area Wilson loops.  Hence at $\beta =0$
\begin{equation}
\langle\overline{\psi}_0\psi_0 \rangle^{{\rm cont\/}}~=~ -2 C K_B~,
{}~~ C\equiv 2^{D/2} N~,
{}~~K_B \equiv {1\over 2 (m_q + D)}~,
\end{equation}
where \hbox{$\psi^{\rm cont\/} = \sqrt{2 K_B}\psi$}
and $C$ is from the Dirac and color traces.  The
$r=1$ pion mass obeys~\cite{wilson}
\begin{equation}
\cosh(M_\pi )   =  1  +
{1-4K_B^2 (D+1) + 16 D K_B^4 \over
8 K_B^2 \bigl(1-2(D-1) K_B^2 \bigr)}~,
\end{equation}
which has $K_c = 1/4$ at $D=4$.

	Upon the approximation
$T_{SA}\to T$ in Eq.~(\ref{eq:branches}),
the sum over trees is implemented by solving the recursion
relation
\begin{equation}\label{eq:noback1}
P_q(x)= - K_B ~\delta_{x,0}
- K_R\sum_{n} f_N(J_n)~ \rlap{$\slash$}{n}
P_q (x-n)~.
\end{equation}
\noindent{}Fourier transform
\hbox{$ P_q (x) \propto \sum_p {\rm e\/}^{ipx} \widetilde{P}_q (p)$}
produces
\begin{equation}\label{k1prop}
\widetilde{P}_q (p) = -{C \over g_0 } ~\Biggl[
{ [ m_R + \sum_\mu \overline{g}_\mu (1-\cos p_\mu ) ]
-i \sum_\mu \gamma_\mu \overline{g}_\mu \sin p_\mu
\over
[ m_R + \sum_\mu \overline{g}_\mu (1-\cos p_\mu ) ]^2
+ \sum_\mu \overline{g}_\mu^2 \sin^2 p_\mu  } \Biggr]~,
\end{equation}
\begin{equation}
g_\mu(\lambda) \equiv f_N(2N\lambda ), ~~~
\overline{g}_\mu \equiv {g_\mu \over g_0} , ~~~
m_R \equiv
{1\over 2K_B g_0} -\sum_{\mu = 0}^{D-1} \overline{g}_\mu .
\end{equation}
For $i\in\{1,\cdots,D-1\}$
$\overline{g}_i$ is given
in Eq.~(\ref{gidefp}).  $M_q^{(\lambda )}$, the
$\vec{p}=0$ pole of $\widetilde{P}_q (p)$,
is related to continuum pole $m_R$ by
\begin{equation}
\sinh{M_q^{(\lambda)} }= m_R~.
\end{equation}

	In Landau gauge where $\overline{g}_\mu = 1$,
${P}_q$ reduces to free particle form with
$m_R = m_q / g_0$.  Otherwise,
${P}_q$ obeys a noncovariant Dirac equation which in
continuum language is
\begin{equation}\label{dirac1}
\bigl( i \sum_{\mu } \overline{g}_\mu \gamma_\mu \partial_\mu
+ M_q^{(\lambda )} \bigr) P_q (x) = 0~.
\end{equation}
Eq.~(\ref{dirac1}) corresponds to
dispersion relation (\ref{dispersion1p}).

	When $K\to K_c = 1/4$, $M_q^{(\lambda )}$ is negative---for
nonnegative values of $M_q^{(\lambda )}$ there is no
massless pion.   This reflects the absence
of chiral symmetry and spontaneous
chiral symmetry breaking at $(r=1,\beta=0)$.  Thus we
turn to
$(r=0,\beta=0)$ in order to examine the $\lambda$-dependence
of $M_q^{(\lambda )}$ in the presence
of spontaneous chiral symmetry
breaking.
\subsection{$N\to\infty$, $r=\beta =0$ Quark Mass}
\label{quarkie}

	When $r=0$, by virtue of \hbox{$\gamma_{n} ~
\gamma_{-n} = 1$}, the Dirac matrix of each
branch is $1$ and the
sum over branches in~(\ref{eq:branches}) is equivalent to
renormalizing $K_B$ to~\cite{patel}
\begin{mathletters}
\begin{equation}\label{eq:renorm1}
K_R ~\equiv~ K_B W(K_B)~.
\end{equation}
$W(K_B)$, the weighted sum over all branch
configurations at one site~\cite{patel,siu}, is deduced as
follows.  The
number $I(L)$ of length $L\ge 1$ ``irreducible'' branches,
branches with a single base stem, is
\begin{equation}\label{eq:recurr}
I(L)= (2D-1)\sum_{p=0}^{L-1}\sum_{ \{l_i \}\atop
\sum_{i=1}^p l_i = L-1 }
\prod_{i=1}^p\Bigl[ {2D-1\over 2D} I(l_i) \Bigr]
{}~~~~ (I(0)=0),
\end{equation}
the sum over all arrangements of its
irreducible subbranches.  At any site
\begin{equation}
W(K_B) = 1+\sum_{L=1}^\infty\bigl(-K_B^2\bigr)^L I(L)
= {1\over 1- w_I (K_B)}
\end{equation}
where
\begin{equation}
w_I (x)   \equiv \sum_{L=1}^\infty x^L I(L)
\label{eq:defwi}
  =  {x\bigl(2D-1\bigr) \over 1- {2D-1\over 2D}w_I(x)}   .
\end{equation}
Definition~(\ref{eq:defwi}) factorizes the RHS of
(\ref{eq:recurr}) to give the RHS of~(\ref{eq:defwi}).  Equating
the RHS
of~(\ref{eq:defwi}) to the LHS gives a polynomial whose
solution yields
\begin{equation}\label{eq:renorm}
K_R~=~1/ \bigl(m_q + \sqrt{ m_q^2 + (2D -1)} \bigr)   .
\end{equation}
\end{mathletters}
Eq.~(\ref{eq:renorm1}) reduces
(\ref{eq:branches}) to
\begin{equation}\label{eq:pullout}
P_q (x) =  \sum_{T_{SA}} \prod_{\{n\}\in T_{SA}}(-K_R)
{}~\rlap{$\slash$}{n}~f_N(J_n)~.
\end{equation}

	In this approach gauge invariant
$(r=0,\beta=0)$ meson propagators
are~\cite{patel,kawamoto}
\begin{equation}
\langle \overline{\psi}_x \Gamma \psi_x
\overline{\psi}_0 \Gamma \psi_0\rangle = N
\sum_{T}{\rm tr\/}\Bigl(\Gamma~
\bigl[ \prod_{\{n\}\in T} K_R \rlap{$\slash$}{n}\bigr]
{}~\Gamma~ \bigl[\prod_{\{m\}\in T^{\dag}}
K_R ~\rlap{$\slash$}{n}\bigr]
\Bigr)
\end{equation}
\noindent{}where $T$ may be nonselfavoiding since
no gaugefixing is involved.  Since trees are
nonbacktracking---a backtrack makes a branch---its Fourier
transform $D_m (p)$ obeys
nonbacktracking randomwalk recursion relation
\begin{equation}\label{eq:split}
{\rm tr\/}[D_m \Gamma ]
= C - K_R^2 \sum_n {\rm tr\/}\bigl[\gamma_m \Gamma \gamma_m
(1-\delta_{n,-m}) D_n {\rm e\/}^{-ipm}\bigr]
\end{equation}
where $\Gamma =\gamma_5$ for the pion and $\gamma_3$ for
the rho.  The commutation of $\Gamma$ with the
$\gamma_m$ splits
\begin{equation}
M_\rho^2 ~=~  4  +  M_\pi^2
\end{equation}
from
\begin{equation}
M_\pi^2 ~=~ {4 (D-1)\over\sqrt{2D -1} }  m_q
+  {\cal O}(m^2)~.
\end{equation}
While $M_\rho$ does not vanish as $m_q\to 0$,
$M_\pi$ vanishes and the pion (actually $2^D$ pions)
plays the role of Goldstone boson for chiral symmetry
breaking.  The same sort of arguments
give~\cite{siu,kawamoto}
\begin{equation} \label{psibarpsib0}
\langle\overline{\psi}_0\psi_0 \rangle  =   -C
{D\sqrt{2 D - 1+m_q^2 }-(D-1)m_q^2 \over
D^2 + m_q^2 }~, ~~~
C = 2^{D/ 2} N
\end{equation}
for the chiral order parameter, which is nonzero when
$m_q\to 0$.

\subsubsection{Nonbacktracking Approximation}

	Since including backtracking
overcounts branches, already summed by $K_R$, we will
enforce nonbacktracking but otherwise
permit recurrence and selfintersection.  The
sum over nonbacktracking trees
is implemented by
\begin{equation}\label{eq:noback}
P_q(x)_n= -{\langle\overline{\psi}_0 \psi_0 \rangle
\over 2D C} ~\delta_{x,0}
- K_R~ f_N(J_n)~ \rlap{$\slash$}{n}
\sum_{m}(1-\delta_{m,-n}) P_q (x-n)_m
\end{equation}
\noindent{}where \hbox{$P_q (x) =\sum_n P_q (x)_n$}.
Subscript $n$ indicates the direction from
which site $x$ was approached.  The Fourier transform
of Eq.~(\ref{eq:noback}) produces
\begin{equation}
\widetilde{P}_q (p)_n = -
{\langle\overline{\psi}_0 \psi_0 \rangle
\over 2D C} +
\sum_m {\cal M\/}_{ n m}(E,\vec{p})  \widetilde{P}_q (p)_m~,
\end{equation}
\begin{equation}
{\cal M\/}_{n m}[E,\vec{p}]   =
 -K_R ~f_N(J_n)~ \rlap{$\slash$}{n}~
(1-\delta_{m,-n}) ~ {\rm e\/}^{ipn}~,
\end{equation}
where ${\cal M\/}$ is a $2D\times 2^{[{ D/ 2}]}$ matrix.

	Effective quark mass $M_q^{(\lambda)}$ obeys
\begin{equation}\label{eq:det}
\det\Bigl(1-{\cal M\/} [
M_q^{(\lambda)} ,\vec{0}]\Bigr) ~=~ 0   .
\end{equation}
Explicit solution by Mathematica$^{\copyright}$ at
$D=2$ and $D=4$ reveals that,
despite being a $2D$ polynomial in
${\rm e\/}^{-M_q^{(\lambda)} t}$, Eq.~(\ref{eq:det})
has only two distinct solutions
corresponding to $\pm M_q^{(\lambda)}$.  The positive energy
solution reduces to
\begin{equation}
\sinh{M_q^{(\lambda)} }=
{1 - (2D-3)\widetilde{g}^2 K_R^2-
g_0^2 K_R^2
- (2D -1) \widetilde{g}^2 g_0^2 K_R^4 \over
2( 1 + \widetilde{g}^2 K_R^2) g_0 K_R }
\end{equation}
where $\widetilde{g}\equiv g_i$ for $i\in\{1,\cdots,D-1\}$.  In
the absence of gaugefixing $g_\mu \to 0$ and
$M_q^{(\lambda)} \to\infty$.  In the absence of renormalization
and perfect gaugefixing,
$K_R\to 1/(2m_q)$,
$g_\mu \to 1$ and the free particle relation
$\sinh{M_q^{(\lambda)} }=m_q$ is recovered.

	Following Eq.~(\ref{eq:linep}), the chiral
limit is extracted
by replacing $K_R$ with $m_q$ according to Eq.~(\ref{eq:renorm})
and expanding about $m_q =0$.  If \hbox{$d\equiv 2D-1$},
\begin{mathletters}
\begin{equation}
\label{eq:mcriticalfull}
M_c^{(\lambda )}   =
  {1 +4D(D-1)
-(3+4D(D-2))\widetilde{g}^2
-d(g_0^2 +\widetilde{g}^2 g_0^2)\over
\sqrt{d} ( 2d + 2\widetilde{g}^2) g_0)} ,
\end{equation}
\begin{equation}
\label{eq:bfull}
B^{(\lambda)}  =
{1 +4D(D-1)+
d (2D\widetilde{g}^2+g_0^2)
-(d-2)\widetilde{g}^4
+(\widetilde{g}^2 + 3d -1)\widetilde{g}^2 g_0^2
\over
2( d + \widetilde{g}^2 )^2 g_0}  .
\end{equation}
\end{mathletters}
\noindent{}The
choice $\lambda_i\equiv \widetilde{\lambda} =1$
(for \hbox{$i=1,\cdots, D-1$})
yields~(\ref{eq:mcd0p}) and
(\ref{eq:bl0p}).  While $M_c^{(\lambda )}$ is continuous,
its second derivative is discontinuous at $\lambda =1/2$.

\subsubsection{$D\to\infty$ Limit}

	In the $D\to\infty$ limit, there is no difference
between selfavoiding and random paths.  In this limit,
the leading ${\cal O}({1/ D})$, $r=0$ effective quark mass
is straightforwardly expressible in terms of the chiral symmetry parameter:
\begin{equation}\label{eq:mcd}
M_c^{(\lambda )}~=~ {7 C\over 16 \langle
\overline{\psi}_0\psi_0 \rangle ~g_0 }
,~~~~
B^{(\lambda)} ~=~ {25 \over 32 g_0 }.
\end{equation}
$1/g_0(\lambda) \equiv 1/f_N(2N\lambda )$ is
a continuous, monotonically decreasing function
whose $2^{\rm nd}$ derivative is discontinuous at $\lambda =1/2$.

\subsection{Matching Coefficients}
\label{sec:matchingsc}

	Call the weighted sum of (naive) lattice
operators whose matrix elements reproduce matrix elements
of a continuum QCD operator the ``lattice representation''
of said continuum operator and the weights ``lattice matching
coefficients.''  Verifying
the $\Delta I=1/2$ Rule in lattice gauge theory is a longstanding
unsolved problem because it has not been possible to
determine all the lattice
matching coefficients of the continuum electroweak
Hamiltonian responsible for $K\to\pi$ matrix
elements, related to $K_s^0\to\pi^+\pi^-$ matrix
elements by chiral perturbation theory.  Specifically,
we are interested in operators~\cite{bs}
\begin{mathletters}
\begin{equation}\label{eq:expanp}
{O_\pm}^{\rm cont\/}[LL] \equiv
z^J_{\pm} J^{\rm latt\/}+
\sum_{\Gamma_1\Gamma_2}
z^{\Gamma_1\Gamma_2 }_\pm
{O_\pm}^{\rm latt\/} [\Gamma_1\Gamma_2 ]~,
\end{equation}
where
\begin{equation}
O_\pm^{qq} [\Gamma_1\Gamma_2] \equiv F^{qq}_{\Gamma_1\Gamma_2}
\pm H^{qq}_{\Gamma_1\Gamma_2} ,
\end{equation}
\begin{equation}\label{fhdef}
F^{qq}_{\Gamma_1\Gamma_2}=
\bar{s}_0\Gamma_1 d_0\cdot\bar{q}_0 \Gamma_2 q_0, ~~~
H^{qq}_{\Gamma_1\Gamma_2}=
\bar{s}_0 \Gamma_1 q_0\cdot\bar{q}_0 \Gamma_2 d_0
\end{equation}
\begin{equation}
\Gamma_1\Gamma_2 \in\{
LL,SS,PP,TT,LR\}~, ~~~~
J(0)\equiv  \bar{s} (0) d(0)~.
\end{equation}
\end{mathletters}
The RHS of (\ref{eq:expanp}) is the combination
of lattice operators
required to reproduce $K\to\pi$ matrix
elements of continuum operator
\hbox{${O_\pm}^{\rm cont\/}[LL]$}.  ``Matching'' coefficients
$z^{\Gamma_1\Gamma_2}_\pm$ are determined by gauge invariant
numerical methods.  The
problem at hand is determining $z_\pm^{J}$.  Proportional
to $(m_s +m_d) a^{-2}$, the $z^J_{\pm}$ coefficients
are beyond weak coupling perturbation theory or
the usual lattice methods.  As described in Ref.~\cite{bs},
we use lattice gaugefixing as a technical device to
determine values of these
matching coefficients.  The idea is to
nonperturbatively replicate what is done in WCPTh, that is, to
make the lattice representations of the continuum operators
reproduce an ansatz, motivated by
flavor symmetry considerations, for
continuum quark correlation functions.  This gauge
covariant matching
condition imposes constraints on the lattice matching coefficients
sufficient to determine them.  We find
that matching coefficients numerically determined in this way are
$\lambda$-independent (within jackknife errors)---as they
must be if they ultimately contribute to the matrix elements of
gauge invariant continuum operators.

	The matching condition invoked in Ref.~\cite{bs}
\begin{equation}
\langle s\vert{O_\pm}^{\rm cont} [LL ]\vert d\rangle~=~0~,
\end{equation}
the parity even quark-equivalent of the
parity odd hadronic renormalization condition that
\hbox{$K ~{\rlap{\slash}{\to}}~ {\rm vac\/}$}, implies that
\begin{mathletters}
\begin{equation} \label{justify}
z^J_\pm~=  -
\sum_{\Gamma_1\Gamma_2  }
\alpha_{{O_\pm} [\Gamma_1\Gamma_2 ]} z^{\Gamma_1\Gamma_2}_\pm ~,
\end{equation}
\begin{equation} \label{eq:a1p}
\langle s\vert {O_\pm}^{\rm latt\/}
[\Gamma_1\Gamma_2 ]\vert d\rangle
{}~\equiv~
\alpha_{{O_\pm} [\Gamma_1\Gamma_2]} \langle s\vert J^{\rm latt\/}
(0)\vert d\rangle  .
\end{equation}
\end{mathletters}
Since quarks are not part of the physical
$S$-matrix, Eq.~(\ref{eq:a1p}) may give a value of
\hbox{$\alpha_{{O_\pm} [\Gamma_1\Gamma_2]}$}
different from its physical sector value.  In the Schwinger
model,~\cite{thesis}
not all matching conditions can be satisfied.  For
the satisfiable ones, different gauge variant matching
conditions (relations between different
gauge variant correlation functions)
may lead to different (or similar) values for matching
coefficients.  The differences stem from
unphysical gauge variant
modes due to gaugefixing ambiguities---gauge
variant operators
transform differently under
residual gauge symmetries.  Nonetheless
matching coefficients are
$\xi$ independent and, thus, plausibly gauge invariant.  Matching
with different
gauge variant correlation functions correspond to
adopting {\it physically\/} inequivalent definitions of the
matching coefficients.  Hence the use of quark matrix elements
(as opposed to, for example, diquark matrix elements) must be
justified in a physical way before proceeding---as done
above Eq.~(\ref{justify}) for this $\Delta I=1/2$ Rule example.

	We have two purposes in this Section:
(a)to understand how $\lambda$-independence of
\hbox{$\alpha_{{O_\pm} [\Gamma_1\Gamma_2]}$} comes about; and
(b)to derive a consequence of $\beta=0$ factorization which
can be compared to
$\beta=5.7$ numerical results.  This comparison
lends insight into why $\beta=0$ formulas mimic so
closely the $\beta =5.7$ and $6.0$ data.

	Following~(\ref{eq:a1p}) we are interested in
$\lambda$-gauge quark correlation functions
of ${O_\pm}^{\rm latt\/} [\Gamma_1
\Gamma_2 ]$ and $J^{\rm latt\/}(0)$.  Because of
$\beta=0$ factorization, quark propagators
in the fermionic Wick expansion of
correlation functions do not interfere with each
other.  Hence
the $\beta=0$ correlation functions are given by replacing
the background field quark propagators in their Wick
expansion with the $\beta=0$ value of
\hbox{$\langle V_x\psi_x\overline{\psi}_0V^{\dag}_0\rangle$}.

	Let $\{a,b\}$ be color and $\{\alpha,\beta\}$
be Dirac indices and define
\begin{equation}
Q^{a b}(t) \equiv
\sum_{\vec{x}} \langle {(V\psi)}^a_x
(\overline{\psi}V^{\dag})^b_0 \rangle, ~~~
\chi^{a b}_{\alpha\beta} \equiv
\langle {(\psi_0)}^a_{\alpha}
(\overline{\psi}_0)^b_\beta\rangle
\end{equation}
where \hbox{$Q\in\{D,S,C\}$} corresponding to
\hbox{$\psi\in\{d,s,c\}$}.  If \hbox{$t_y=-t$},
by (\ref{eq:pullout})
\begin{equation}
\sum_{\vec{y}} \langle{ V_0\psi}_0
\overline{\psi}_y V_y^{\dag}\rangle
{}~=~ \gamma^5 ~Q^{\dag}(-t)~ \gamma^5~.
\end{equation}
By $R$ symmetry and (\ref{psibarpsib0}),
\begin{equation}\label{diagchi}
\chi_{\alpha\beta}^{ab}~=~ -
\delta^{ab}~\delta_{\alpha \beta}~
\langle\overline{\psi}_0\psi_0\rangle / (2^{D/2} N) ~.
\end{equation}

	In fact, a stronger statement can
be made for $\chi$.  Let ``$\beta=0$
graph'' refer to any single
hopping expansion graph of
$[\psi_0\overline\psi_0]_f$ which contributes nontrivially
to $\chi$ at $\beta=0$.  Then, since whenever
$\bigcirc_x$ encloses zero area
\hbox{$\prod_{\{y,n\}\in \bigcirc_x }
U_{y,n}~\rlap{$\slash$}{n} = 1$}
and the $[dU]$
integral in~(\ref{diagchi}) is trivial,
\begin{equation}
{[(\psi_0)^a_\alpha (\overline{\psi}_0)^b_\beta ]_f}_{\vert
\beta=0~{\rm graph\/}} ~
\propto~ \delta^{ab}  \delta_{\alpha\beta} ~.
\end{equation}
This color diagonality of
$[\psi_0\overline{\psi}_0]_f$ permits
factorization of correlation functions containing
$[\psi_0\overline{\psi}_0]_f$
``bubble'' contractions at $\beta=0$.  While
$[\psi_0\overline{\psi}_0]_f$
graphs are not diagonal at $\beta >0$, at
$\beta=5.7$
\begin{equation}\label{smallflucts}
[(\psi_0)^a_\alpha (\overline{\psi}_0)^b_\beta ]_f
{}~\propto~ \delta^{ab}\delta_{\alpha\beta}
{}~+~ 0.1\% ~ {\rm fluctuations\/}.
\end{equation}
On a typical $16by24$ gauge configuration with
$\lambda=1$, $K=.094$, and in Dirac space
\begin{equation}\label{eq:mat1}
{[\psi_0^{1} \overline{\psi}_0^{1}]_f} =
{2 K\over 10^3 }
\pmatrix{
(995.,.1)&(.0,.0)&(.0,.0)&(.1,
.0)\cr
(.0,.0)&(995.,.1)&(.1,.0)&(.0,
.0)\cr
(-.2,.0)&-(.1,.3)&(995.,-.1)&(.0,
.0)\cr
(-.1,.3)&(.2,.0)&(.0,.0)&(995.,
-.1)\cr};
\end{equation}
with $K=.166$,
\begin{equation}\label{eq:mat2}
{[\psi_0^{1} \overline{\psi}_0^{1}]_f} =
{2 K \over 10^3} \pmatrix{
(902.,.9)&-(1.1,
.6)&(-.4,.0)&(-.9,2.1)\cr
(2.8,.9)&(903.,4.2)&-(.9,
2.1)&(-1.9,.0)\cr
(-2.7,.0)&-(1.1,4.1)&(902.,-.9)&(2.8,
-.9)\cr
(-1.1,4.1)&(.7,.0)&(-1.1,.6)&(903.,
-4.2)\cr} .
\end{equation}
In the color off-diagonal sector with $K=.166$,
\begin{equation}\label{eq:mat3}
{[\psi_0^{1} \overline{\psi}_0^{2}]_f} =
{2K\over 10^3} \pmatrix{
-(8.0,3.6)&(5.7,2.1)&(11.,5.1)&(.04,
8.4)\cr
(2.6,-2.3)&-(9.8,7.4)&-(20.,8.4)&-(5.3,
8.2)\cr
-(2.1,7.0)&(6.5,1.1)&(4.0,.00)&(7.3,
5.1)\cr
(5.9,7.6)&(1.4,11.)&(3.9,-1.2)&(1.3,
-13.)\cr}  .
\end{equation}
In general, $(I)$fluctuations grow as $K$ increases;
$(II)$in the color singlet sector, scalar components dominate;
$(III)$in the color nonsinglet sector, the
{\it non\/}scalar fluctuations
are greater than the scalar, sometimes by as much as
an order of magnitude;
$(IV)$these statements are valid for all $\lambda$
and axial gauge configurations with typical
variation between configurations of $\sim 20\%$.

        The Euclidean Dirac matrices we use
obey Fierz relations
\begin{mathletters}
\begin{equation}\label{fierzrelationp}
\Gamma_1^{\alpha\beta}\Gamma_2^{\gamma\delta}
{}~=~ \sum_{\Gamma_3\Gamma_4} f_{\Gamma_1\Gamma_2}^{
\Gamma_3\Gamma_4} ~ \Gamma_3^{\alpha\delta} \Gamma_4^{\gamma\beta}
\end{equation}
with
\begin{equation}\label{fierzratsp}
f^{SS}_{SS}= f^{SS}_{PP} = {1\over 4}~, ~~
f^{SS}_{VV} = - f^{SS}_{AA} = 1~, ~~
f^{SS}_{TT} = 3~,
\end{equation}
\begin{equation}\label{eq:ff1p}
[\gamma^\mu (1\pm\gamma^5)]_{ij}
[\gamma^\mu (1\pm\gamma^5)]_{kl} =
-[\gamma^\mu (1\pm\gamma^5)]_{il}
[\gamma^\mu (1\pm\gamma^5)]_{kj}.
\end{equation}
\end{mathletters}
If  \hbox{$t_y = -t_x \equiv -t$}, Wick expansion and
Eqs.~(\ref{fhdef}), (\ref{diagchi}) and
(\ref{fierzrelationp}-\ref{eq:ff1p}) imply
\begin{equation}\label{jmatelem}
{\cal J\/}\equiv
\sum_{\vec{x},\vec{y}} \langle V_x s_x~J(0)~\overline{d}_y
V_y^{\dag} \rangle =
S(t)~ \gamma_5~D^{\dag}(-t)~ \gamma^5~,
\end{equation}
\begin{equation}\label{lmatelem}
\sum_{\vec{x},
\vec{y}} \langle V_x s^c_x H^{qq}_{\Gamma_1\Gamma_2}\overline{d}^d_y
V_y^{\dag} \rangle =\cases{
{}~~\sim {\rm tr\/}(
\chi^{ab} L)= 0 & $\Gamma_1\Gamma_2 =LL$;\cr
-{1\over 2^{D/2} N}f^{SS}_{\Gamma_1\Gamma_2}~{\cal J\/}^{cd}
\langle\overline{\psi_0}\psi_0\rangle &otherwise;\cr}
\end{equation}
\begin{equation}\label{fmatelem}
\sum_{\vec{x},\vec{y}}
\langle V_x s^c_x F^{qq}_{\Gamma_1\Gamma_2}\overline{d}^d_y
V_y^{\dag} \rangle =
\cases{ {\cal J}^{cd}~\langle
\overline{\psi}_0\psi_0\rangle  &
$\Gamma_1\Gamma_2 = SS$;\cr
{}~~~~0~~~~ & otherwise.\cr}
\end{equation}

	Comparing (\ref{lmatelem}) and (\ref{fmatelem})
to (\ref{jmatelem}) yields
\begin{equation}
\alpha_{F^{qq}_{\Gamma_1\Gamma_2}}~=~\cases{
\langle\overline{\psi}_0\psi_0\rangle  &
$\Gamma_1\Gamma_2 = SS$;\cr
{}~~~0~~~ & otherwise;\cr}
\end{equation}
\begin{equation}
\alpha_{H^{qq}_{\Gamma_1\Gamma_2}}~=~\cases{ ~~~~0~~~~ & $LL$;\cr
-{1\over 2^{D/2} N}~\langle\overline{\psi_0}\psi_0\rangle ~
f^{SS}_{\Gamma_1\Gamma_2} &otherwise.\cr }
\end{equation}
Therefore, $\alpha_{H^{qq}_{\Gamma_1\Gamma_2}}$
is the dominant contribution to
$\alpha_{{O_\pm}^{qq}[\Gamma_1\Gamma_2]}$
unless \hbox{$\Gamma_1\Gamma_2=SS$}, in which case
$\alpha_{H^{qq}_{SS}}$ is $1/N$ suppressed
relative to $\alpha_{F^{qq}[SS]}$.  By
(\ref{fierzratsp}) the $\beta=0$ ratios are
\begin{equation}
\alpha_{H^{qq}_{SS}}:\alpha_{H^{qq}_{PP}}:
\alpha_{H^{qq}_{VV}}:\alpha_{H^{qq}_{AA}}:
\alpha_{H^{qq}_{TT}}
 ~=~1 :  1  :   4   :  -4  :   12~,
\end{equation}
which as described in Ref.~\cite{bs}
are approximately numerically reproduced at
$\beta=5.7$ and $6.0$.  Furthermore,
appropriate linear combinations of the $\alpha_\Theta$ give
Eq.~(\ref{alphab0valp}).  Since \hbox{$\langle \overline{\psi}_0
\psi_0\rangle$} is gauge invariant, \hbox{$\alpha_{O_\pm [
\Gamma_1\Gamma_2 ]}$} is gauge invariant.

\acknowledgments

	I am indebted to Amarjit Soni and Claude Bernard for
inspiring my interest in gaugefixing.  Special credit is due
Claude Bernard for
calling my attention to several key issues, including
recurrent quark paths and how gaugefixing induces Wilson lines
between sources, and for
keeping me honest.  This project
would not and could not have been pursued without Bernard's and
Soni's support.  The computing was done at
the National Energy Research Supercomputer Center (partially within
``Grand Challenge'').  This manuscript has been authored under
contract $\#DE-AC02-76CH00016$ with the U.S.~D.O.E.

\end{document}